\documentclass[traditabstract]{aa}
\usepackage{natbib}
\usepackage{epsfig}
\usepackage{txfonts}
\usepackage{graphicx}

\def\ms{\,m\,s$^{-1}$}         %m.s -1
\def\kms{\,km\,s$^{-1}$}         %m.s -1
\def\ms{\hbox{\,m\,s$^{-1}$}}         %m.s -1
\def\m2s2{\hbox{\,m$^{2}$\,s$^{-2}$}} %m2.s -2
\def\kms{\hbox{\,km\,s$^{-1}$}}       %km.s -1
\def\vsini{\hbox{$\upsilon \sin i_{\star}\;$}}      %vsini
\def\Msun{\hbox{$\mathrm{M}_{\odot}~$}}             %Msun
\def\Rsun{\hbox{$\mathrm{R}_{\odot}~$}}
\def\Mjup{\hbox{$\mathrm{M}_{\rm Jup}$~}}
\def\Rjup{\hbox{$\mathrm{R}_{\rm Jup}$~}}

\def\teff{T$_{\rm eff}~$}
\def\logg{log {\it g}}
\def\met{[Fe/H]~}
\def\mr{$M_\star^{1/3}/R_\star$}

\bibpunct{(}{)}{;}{author-year}{}{,}

\begin{document}
\title{\textsc{SOPHIE} velocimetry of \textit{Kepler} transit candidates \thanks{Based on observations made with SOPHIE on the 1.93-m telescope at Observatoire de Haute-Provence (CNRS), France}}
\subtitle{IV. KOI-196b: a non-inflated hot-Jupiter with a high albedo}

\author{
 A. Santerne \inst{1} 
\and A.~S. Bonomo \inst{1}
\and G. H\'ebrard \inst{2,3}
\and M. Deleuil \inst{1}
\and C. Moutou \inst{1}
\and J.-M. Almenara \inst{1}
\and F. Bouchy \inst{2, 3}
\and R.~F. D\'iaz \inst{2,3}
}
\institute{
Laboratoire d'Astrophysique de Marseille, Universit\'e d'Aix-Marseille \& CNRS, 38 rue Fr\'ed\'eric Joliot-Curie, 13388 Marseille cedex 13, France
\and Institut d'Astrophysique de Paris, UMR7095 CNRS, Universit\'e Pierre \& Marie Curie, 98bis boulevard Arago, 75014 Paris, France
\and Observatoire de Haute-Provence, Universit\'e d'Aix-Marseille \& CNRS, 04870 Saint Michel l'Observatoire, France}
\date{Received: 2 August 2011 ; Accepted: 12 September 2011}

\offprints{Alexandre~Santerne\\
 \email{alexandre.santerne@oamp.fr}}

\abstract{We report the discovery of a new hot-Jupiter, KOI-196b, transiting a solar-type star with an orbital period of 1.855558 days $\pm$ 0.6s thanks to public photometric data from the \textit{Kepler} space mission and new radial velocity observations obtained by the SOPHIE spectrograph mounted on the 1.93-m telescope at the Observatoire de Haute-Provence, France. The planet KOI-196b, with a radius of $0.841 \pm 0.032 $ \Rjup and a mass of $0.49 \pm 0.09$  \Mjup, orbits a G2V star with $R_{\star}$ = 0.996 $\pm$ 0.032 \Rsun, $M_\star$= $0.94 \pm 0.09 $ \Msun, \met = $-0.10\pm0.16$ dex,  \teff= 5660 $\pm$ 100 K and an age of $ 7.7 \pm 3.4$ Gy. KOI-196b is one the rare close-in hot-Jupiters with a radius smaller than Jupiter suggesting a non-inflated planet. The high precision of the \textit{Kepler} photometry permits us to detect the secondary transit with a depth of $64 ^{_{+10}}_{^{-12}} $ ppm as well as the optical phase variation. We find a geometric albedo of $A_{g} = 0.30 \pm 0.08$ which is higher than most of the transiting hot-Jupiters with a measured $A_{g}$. Assuming no heat recirculation, we find a day-side temperature of $T_\mathrm{day} = 1930Ê\pm 80$ K. KOI-196b seems to be one of the rare hot-Jupiters located in the short-period hot-Jupiter desert.

}

%\titlerunning{\textit{SOPHIE} follow-up of \textit{Kepler} planetary transiting candidates I.}
\titlerunning{KOI-196b: a non-inflated hot-Jupiter with a high albedo}
\authorrunning{A.~Santerne}

\maketitle

\section{Introduction}
\label{intro}

Since 2007, thanks to the exoplanet-dedicated space missions \textit{CoRoT} \citep{2006cosp...36.3749B} and \textit{Kepler} \citep{2010Sci...327..977B}, the scientific community has access to very high precision photometry, down to a few tens of ppm with times series up to a few years. This has led to the exciting discoveries of the super-Earths \textit{CoRoT}-7b \citep{2009A&A...506..287L,2009A&A...506..303Q} and its twin \textit{Kepler}-10b \citep{2011ApJ...729...27B}, the multi-planetary systems \textit{Kepler}-9 \citep{2010Sci...330...51H} and \textit{Kepler}-11 \citep{2011Natur.470...53L} and the long-period giant planet \textit{CoRoT}-9b \citep{2010Natur.464..384D}. Long space-based time series allow us to reach a very accurate characterization of the planetary systems, especially when stellar parameters can be determined by asteroseismology \citep[e.g.][]{2011ApJ...729...27B,2011ApJ...735L..12D}. \\

Moreover, high-precision space-based photometry permits us to find also very small effects, such as the planetary occultation \citep[e.g.][]{2009A&A...501L..23A}, i.e. when the planet passes behind its host star, as well as the phase variation of the planet, i.e. the variation in brightness as the dayside of the planet rotates into view. The latter was found in the optical for \textit{CoRoT}-1b \citep{2009Natur.459..543S}, HAT-P-7b \citep{2009Sci...325..709B, 2010ApJ...713L.145W}, \textit{Kepler}-7b \citep{2011ApJ...735L..12D} and the super-Earth \textit{Kepler}-10b \citep{2011ApJ...729...27B}.\\

The first six months of \textit{Kepler} data have permitted identifying 1235 planetary candidates around 997 stars \citep{2011arXiv1102.0541B} including twenty one fully characterized planets (with measured radius and mass) \citep[e.g.][]{2010ApJ...713L.126B, 2010ApJ...713L.140L, 2011A&A...528A..63S, 2011arXiv1106.3225B} and three confirmed planets without measured mass \citep[e.g.][]{2011ApJ...727...24T}. Out of the remaining candidates, we have selected a few candidates around stars brighter than the \textit{Kepler} magnitude $K_p\sim14.7$ to be followed up with the SOPHIE spectrograph (Observatoire de Haute-Provence, France). One of these is the new exoplanet KOI-196b. In this paper we report its discovery (Sect. \ref{keplerobs} \& \ref{sophieobs}) and the system characterization (Sect. \ref{spclass}) including an estimate of the planetary albedo and day-side temperature (Sect. \ref{albedocharac}) thanks to the observation of the secondary and phase variation (Sect. \ref{sect.occultation}). Finally, we discuss KOI-196b and compared it to other transiting planets in terms of mass, radius, period and day-side temperature (Sect. \ref{discussion}).

\section{\textit{Kepler} observations  \label{sect.keplerdata}}
\label{keplerobs}
\begin{figure}[h]
\begin{center}
\includegraphics[width=\columnwidth]{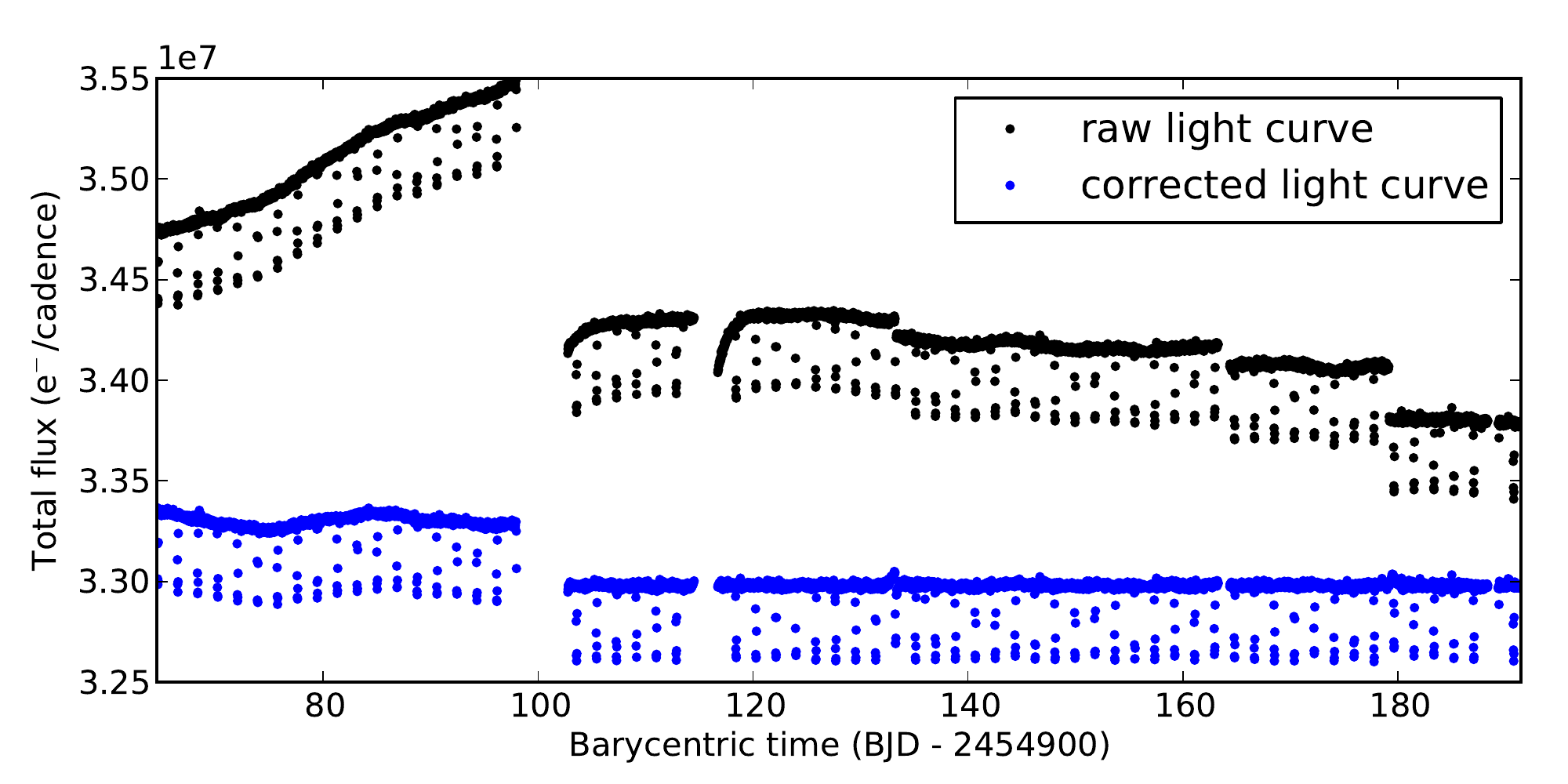}
\caption{Raw and corrected light curve of KOI-196 as provided by the MAST database.}
\label{196rawlc}
\end{center}
%\vspace{-0.5cm}
\end{figure}

The \textit{Kepler} Object of Interest KOI-196 has been observed by \textit{Kepler} since May 13, 2009 in long-cadence mode (temporal sampling of $\sim$ 29.4 min). The various identifiers (ID) of this target, including coordinates and magnitudes are reported in Table \ref{keplerID}. At the time of writing of the present paper, only quarters 1 \& 2 (Q1 \& Q2) data sets are publicly available on the MAST archive\footnote{http://archive.stsci.edu/}. They account for a total of  5~993 photometric measurements. 285 points were discarded by the \textit{Kepler} pipeline \citep{2010ApJ...713L..87J} as they are affected by instrumental effects. The 5708 remaining photometric measurements are displayed in Fig. \ref{196rawlc}. Transits with a period of $\sim$ 1.8 days and a depth of $\sim$ 1 \% are clearly visible in the light-curve of KOI-196 showing no prominent features related to magnetic activity.\\

\begin{table}[]
\centering
\caption{KOI-196 IDs, coordinates and magnitudes}
\renewcommand{\footnoterule}{}                          
\begin{minipage}[t]{7.0cm} 
\begin{tabular}{cc}
\hline
\hline
\textit{Kepler} Input Catalog (KIC) & 9410930 \\
\textit{Kepler} Object of Interest (KOI) & 196.01 \\
2MASS ID & 19380317+4558539 \\
%GSC2.3 & N2HG047856\\
%USNO-A2 & 1350-11065312\\
 \hline
Right Ascension (J2000) & 19 38 03.19 \\
Declinaison (J2000) & 45 58 53.76  \\
\hline
\textit{Kepler} magnitude$^{(a)}$ & 14.465 $\pm$ 0.02\\
 $g^{(a)}$  & 14.996 \\
 $r^{(a)}$  & 14.410 \\
 $i^{(a)}$  & 14.238 \\
 $z^{(a)}$  & 14.182 $\pm$ 0.030 \\
 $J^{(b)}$ & 13.262 $\pm$ 0.022\\
 $H^{(b)}$ & 12.938 $\pm$ 0.024\\
 $K^{(b)}$ & 12.892 $\pm$ 0.028\\
 $E(B-V)^{(a)}$ & 0.114 $\pm$ 0.100\\
\hline
\hline
\end{tabular}
%\footnotetext[1]{from the Tycho-2 catalog, the \textit{Kepler} magnitude ($K_p$) is calculated as follows : $blue = 0.54*B + 0.46*V - 0.07$, $red = -0.44*B + 1.44*V + 0.12$, $color = blue - red$. If $color \leq 0.8$ $K_p = 0.8 * red + 0.2 * blue$ else $K_p = 0.9 * red + 0.1 * blue$}
\vspace{-0.3cm}
\footnotetext{$^{(a)}$ from the \textit{Kepler} Input Catalog.~$^{(b)}$ from 2MASS catalog.}
\label{keplerID}      
\end{minipage}
\end{table}

\subsection{Contamination correction}
\label{cont}

For the following studies, we used the raw light curve (hereafter LC) instead of the PDC-corrected light curve as recommended by the \textit{Kepler} Team\footnote{http://keplergo.arc.nasa.gov/PipelinePDC.shtml}. The flux of several nearby stars contaminates the target one and has to be taken into account when working with the raw LC. We estimated the contamination value by comparing the raw and the PDC-corrected LC following the same methodology as described in \citet{2011arXiv1106.3225B}. We calculated it separately for each quarters. We find a contamination value of  $5.1\% \pm 0.06\%$ for Q1 and $3.3\% \pm 0.04\%$ for Q2 which is in good agreement with the values for seasons 3 and 0, respectively, available on the MAST database (see also Sect. \ref{centroid}). \\

\subsection{Primary transit modeling}
\label{sect.transit}

Before performing the transit modeling, we normalized the transits by fitting a parabola to the $\sim 11$h intervals of the LC before the ingress and after the egress of each transit in order to correct for any local variations. We discarded one of the 63 available transits occuring at BJD$\sim$2455033.3 and is affected by one of the discontinuities seen in the LC (see Fig. \ref{196rawlc}). \\

Transit modeling was performed following the formalism of \citet{2006A&A...450.1231G} fixing the eccentricity to zero (see Sect. \ref{sect.occultation}). 
%We did not performed a combined fit including \textit{Kepler} photometry and SOPHIE radial velocities since we did not want to introduce bias in the transit modeling due to not understood systematics effects present when observing faint stars with the SOPHIE high-efficiency mode \citep{2011MNRAS.tmp..424H}.
The seven free parameters used in the transit modeling were the orbital period $P$, the epoch of the first transit $t_{0}$, the transit duration $T_\mathrm{14}$, the planet to star radii ratio $k=R_p/R_\star$, the orbital inclination $I$, and the two limb-darkening coefficients $u_+=u_a+u_b$ and $u_-=u_a-u_b$. \\

\begin{figure}[h]
\begin{center}
\includegraphics[width=\columnwidth]{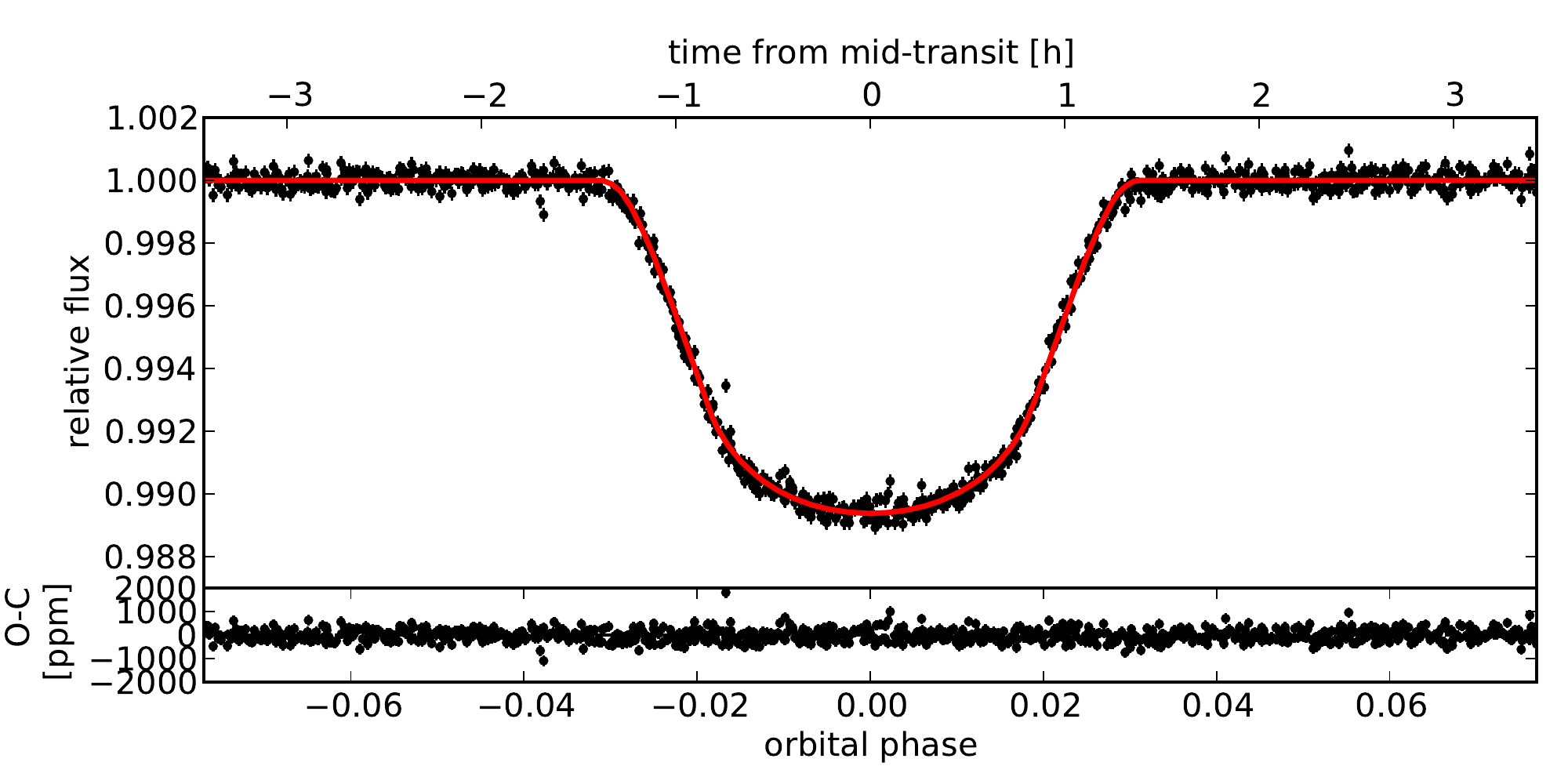}
\caption{Phase-folded unbinned light curve of the KOI-196b transit with its best-fit model (top panel) and residuals (bottom panel).}
\label{196tr}
\end{center}
%\vspace{-0.5cm}
\end{figure}

The best-fit of the primary transit was performed using the algorithm AMOEBA \citep{1992nrfa.book.....P} and changing the initial parameters with a Monte Carlo to find the global minimum of the $\chi^{2}$. We followed the procedure described in \citet{2010MNRAS.408.1758K} to take the long cadence rate of the \textit{Kepler} LC into account: the $\chi^{2}$ was computed by binning a five-times oversampled model LC within the long cadence rate of \textit{Kepler} and compared to the data. The phase-folded transit and the best-fit model as well as the residuals between the observations and the model are shown in Fig. \ref{196tr}. We used a bootstrap procedure to estimate the uncertainties which consist in generating synthetic data sets shifting the residuals from the best-fit, adding them back to the best-fit model LC and fitting once again the data. We finally determined the 1-$\sigma$ uncertainty as the 68\% confidence interval defined as 16\% above the upper (and below the lower) confidence limit of the cumulative probability. The parameters of the best-fit model and their 1-$\sigma$ uncertainties are reported in Table \ref{starplanetparam}.\\

\subsection{Modeling the occultation and orbital phase variations}
\label{sect.occultation}

\begin{figure*}[]
\begin{center}
\includegraphics[width=0.8\textwidth]{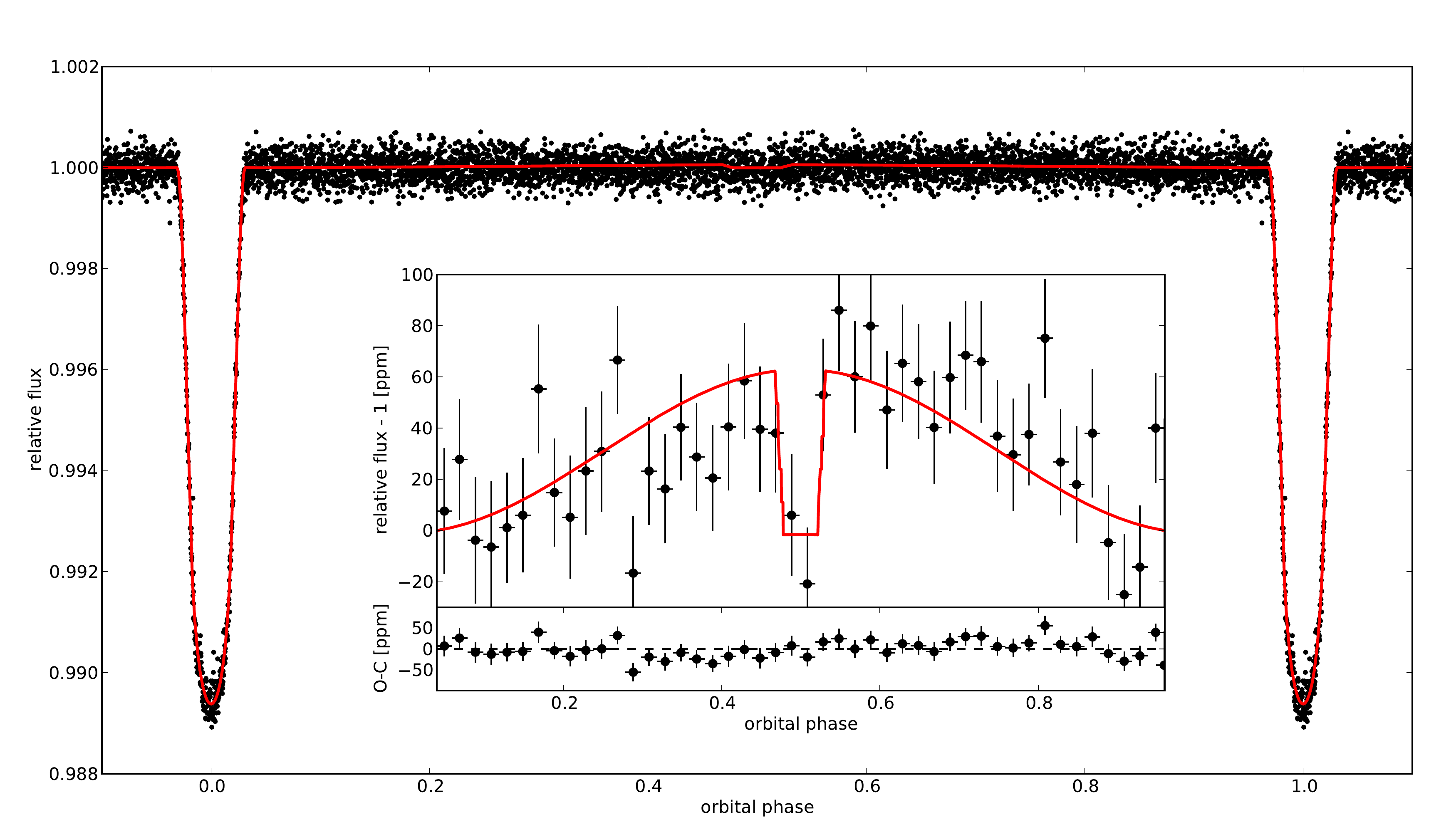}
\caption{Unbinned phase-folded light curve of KOI-196 with the best-fit model. The inset displays a zoom of the out-of-transit LC binned to 0.02 in phase with the best-fit model of the planet phase variation and occultation (inset, top panel) and its residuals (inset, bottom panel). Out-of-transit residuals binned to 0.02 in phase have a dispersion of 22 ppm.}
\label{196lc}
\end{center}
%\vspace{-0.5cm}
\end{figure*}

Space-based high-precision photometry has shown its capability to detect planetary occultation and phase variation for close-in planets \citep[e.g.][]{2009A&A...501L..23A, 2009A&A...506..353A, 2009Natur.459..543S, 2010A&A...513A..76S, 2009Sci...325..709B, 2011ApJ...735L..12D}. Even if KOI-196 is faint, we cleaned the LC to try to detect both effects. To that purpose, we first removed all the transits from the raw LC corrected from the contamination (Sect. \ref{cont}). We then corrected for long-term trends separately in the Q1 and the two first subsets of the Q2 data between BJD 2455000.0 and 2455033.3, by fitting polynomials of the 8$^{th}$ order. For the other subsets of the Q2 LC, we just removed the offsets around each discontinuity. We finally de-trended the whole LC using a sliding median with a window of 1.5 times the orbital period, taking care of mirroring the LC at the beginning and the end. This method leads to a total dispersion of 243ppm after a 3-$\sigma$ clipping. The obtained LC, phase-folded at the transit ephemeris is displayed in Fig. \ref{196lc} and clearly shows an occultation as well as orbital phase variations. We also tested different sliding median windows between $\sim 1.5 \cdot P $ and $\sim 2.5 \cdot P $ and got the same result with a slightly larger dispersion. As expected, windows with extensions less than the orbital period significantly reduce the amplitude of the phase variations we are searching for and thus, our choice of  $1.5 \cdot P $ is the best compromise between preserving the signal we want to detect and minimizing the RMS of the residuals.\\

%We also reduced the LC using a different approach. This second filtering consists in splitting the LC in subsets of $\sim$ 7 orbital period, de-trending them using 2$^{nd}$ order polynomials and sets of sinusoid fitted to data with periods of more than twice the orbital one. We corrected subsets of the LC until obtaining a dispersion at the same level than the mean photometric uncertainty $\left<\sigma_\mathrm{phot}\right>$ = 223ppm. This method is based on the local correction of both stellar variations or instrumental systematics. We finally obtained a dispersion of 244ppm after a 3-$\sigma$ clipping and a equivalent light-curve as the one displayed in Fig. \ref{196lc}.\\

%\begin{table*}[p]
%\centering
%\caption{TBD}
%\renewcommand{\footnoterule}{}                          
%\begin{minipage}[t]{\textwidth} 
%\setlength{\tabcolsep}{1.0mm}
%\begin{tabular}{cc}
%\hline
%\hline
%subsets  & corrections\\
%(BJD-2454900) & \\
%\hline
%\multicolumn{2}{c}{Q1 data}\\
%\hline
%$<$ 100.0 & $  - 37.5523 + 0.0007 \cdot BJD$\\
%& $1.0001+ 0.0011\sin((BJD-22.4792)/22.8180)$\\
% & \\
%\hline
%$<$ 66.5 & $  21.0365 -0.0004\cdot BJD$\\
% & \\
%\hline
%66.5 -- 72.0 & $ 1.0000 + 0.0002\sin((BJD-69.3869)/3.0827)$\\
% & \\
%\hline
%72.0 -- 85.0 & $ -0.3023 + 0.00002 \cdot BJD$\\
% & \\
%\hline
%85.0 -- 85.0 & $ -0.3023 + 0.00002 \cdot BJD$\\
% & \\
%\hline
%\hline
%\end{tabular}
%\vspace{-0.3cm}
%\label{second correction}      
%\end{minipage}
%\end{table*}

We characterized the occultation and phase variations using the formalism proposed by \citet{2009Natur.459..543S}. To model the occultation, we fixed the values of the orbital period, the epoch of transit, the transit duration, the radii ratio and the inclination to those derived by the transit modeling but without introducing the LD coefficient, as the planet is passing behind the star. We first binned the LC to 0.02 in phase (inset of Fig. \ref{196lc}) and fitted an out-of-transit model with four free parameters: the phase of the secondary eclipse $\phi_\mathrm{occ}$, the contrast between the planet day-side and stellar flux $R_\mathrm{day}$, the ratio of the night-side to day-side flux $F_\mathrm{N/D}$ and the relative stellar brightness $z_\mathrm{lev}$ \citep{2009Natur.459..543S}. As for the transit modeling, we found the best-fit solution using the algorithm AMOEBA and changing the initial parameters with a Monte Carlo method to determine the global minimum of the $\chi^{2}$. We also oversampled the model to take the bin size into account. We found the phase of the secondary eclipse to be $\phi_\mathrm{occ} = 0.499 ^{_{+0.004}}_{^{-0.011}}$ which allow us to constrain $e\cos \omega = -0.002 ^{_{+0.006}}_{^{-0.017}}$ \citep{2009ASPC..404..291G}. Thus, we assumed that the eccentricity of KOI-196b is zero. Then, we performed again an out-of-transit modeling to the unbinned LC, fixing the phase of the occultation to 0.5 and allowing the three remaining parameters to vary. We computed the  $\chi^{2}$ by comparing the data with a five-times oversampled model binned at the \textit{Kepler} cadence rate. Indeed, we expected that not oversampling the model could lead to an underestimation of the occultation depth. The best fit corresponding to the global minimum of the $\chi^{2}$ is plotted in Fig. \ref{196lc} and the derived parameters are displayed in Table \ref{starplanetparam}. One-$\sigma$ uncertainties were determined using a bootstrap procedure as for the primary transit (see Sect. \ref{sect.transit}).\\

\subsection{Centroid motion}
\label{centroid}

\begin{figure}[]
\begin{center}
\includegraphics[width=\columnwidth]{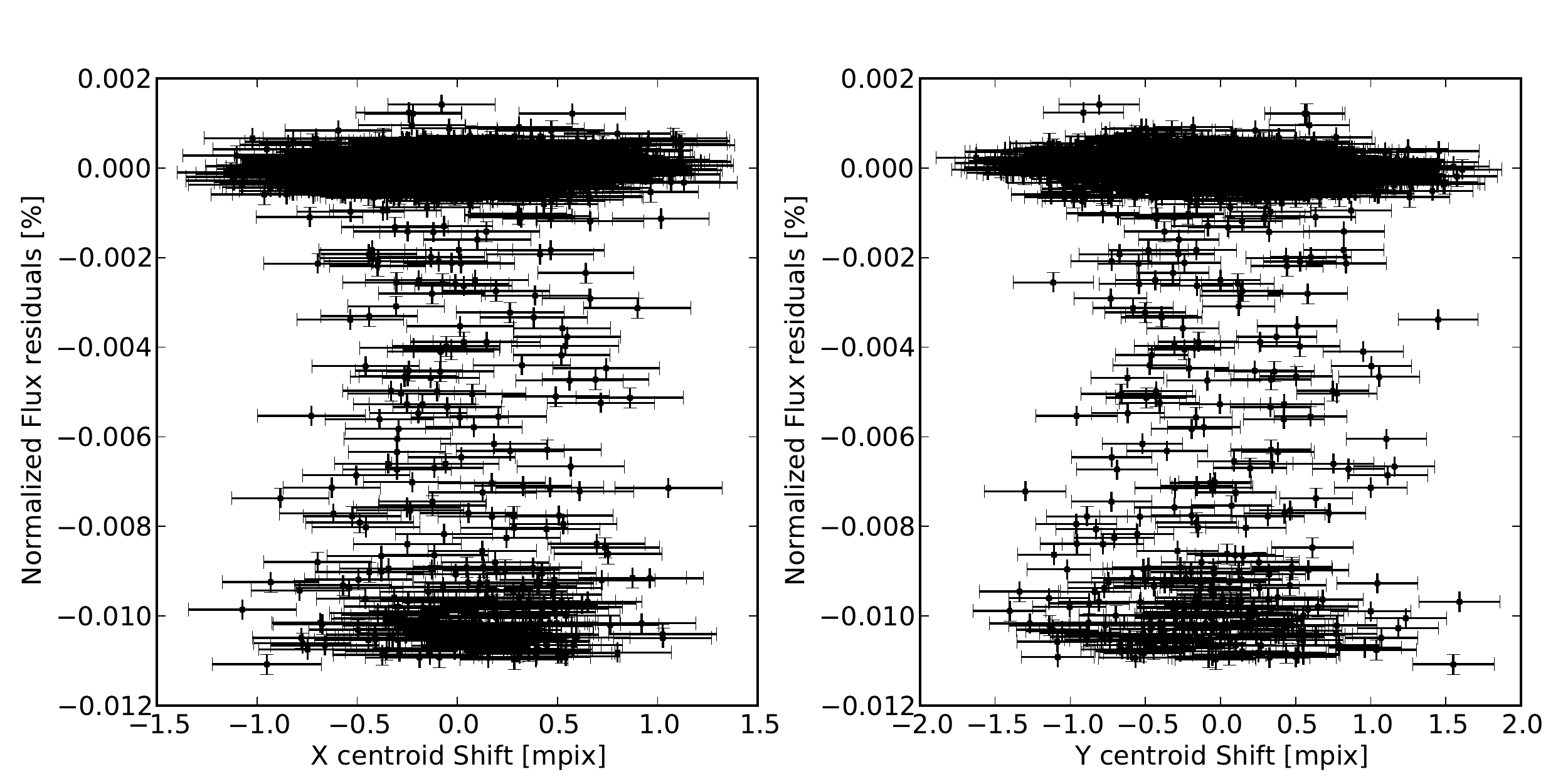}
\caption{Rain plots showing X (left pannel) and Y (right pannel) shift of the centroid during the transit as function of the normalized flux residuals.}
\label{196centroid}
\end{center}
%\vspace{-0.5cm}
\end{figure}

In order to test whether the centroid behaviour was compatible with the target's vicinity \citep{2010ApJ...713L.103B}, we performed simple 2D simulations of the photometric properties of KOI-196 and its five neighbours located less than 0.3\arcmin~ from KOI-196 and described in the MAST archive. All of them are fainter than $K_{p}=16.2$. We used circular apertures instead of the actual \textit{Kepler} PSF; this assumption should impact only the farthest contaminants. It results that all configurations where one of the neighbours produces the transit are discarded, since this would imply a shift in the X and/or Y directions of 2 to 20 mpixels in absolute value. In our simulations, when KOI-196 is the transiting star, we still find a little centroid shift of 1-2 mpixels. The observed centroid positions have a mean value of 0 in both directions, with 1-$\sigma$ errors of 0.4 and 0.6 mpix in X and Y, respectively (see Fig \ref{196centroid}). The simple simulations of the photocenter of KOI-196 being disturbed by its neighbours are therefore in agreement with the observations, within 2-$\sigma$.

%The centroid motion of KOI-196 does not show any shift during the transit (see Fig. \ref{196centroid}). So we can exclude that the two contaminating stars KIC 9410927 ($K_{p}=18.3$) located at 0.1\arcmin and KIC 9410924 ($K_{p}=16.3$) located at 0.14\arcmin are eclipsing binaries with the considered ephemeris.

\section{SOPHIE observations}
\label{sophieobs}

\subsection{Observations and data reduction}

We performed spectroscopic follow-up observations on the target KOI-196 with the SOPHIE spectrograph \citep{2008SPIE.7014E..17P, 2009A&A...505..853B} mounted on the 1.93-m telescope at Observatoire de Haute-Provence, France. We acquired twelve high-resolution spectra from March 25 to July 29, 2011\footnote{prog. ID: 11A.PNP.MOUT} using the High Efficiency mode ($R \sim 39~000$ at 550nm) of SOPHIE and the slow CCD read-out mode in order to minimize the instrumental and photon noise. Observations were done by keeping the SNR of the spectra constant in order to minimize the charge transfer inefficiency effect \citep{2009EAS....37..247B}. Spectra were reduced with the online standard pipeline and radial velocities were obtained by computing the weighted cross-correlation function (CCF) of the spectra with a numerical spectral mask of a G2V star \citep{1996A&AS..119..373B, 2002A&A...388..632P}. Three of the spectra were significantly affected by the Moon's scattered light. Their corresponding radial velocity were corrected using the same technique as in \citet{2011A&A...528A..63S} and \citet{2010arXiv1006.2949B}. Other two spectra were slightly affected by the Moon scattered light and were not corrected since the Moon light affected the measurement at the level of a few \ms, which is compatible with the noise added by the correction. \\

\begin{figure}[h]
\begin{center}
\includegraphics[width=\columnwidth]{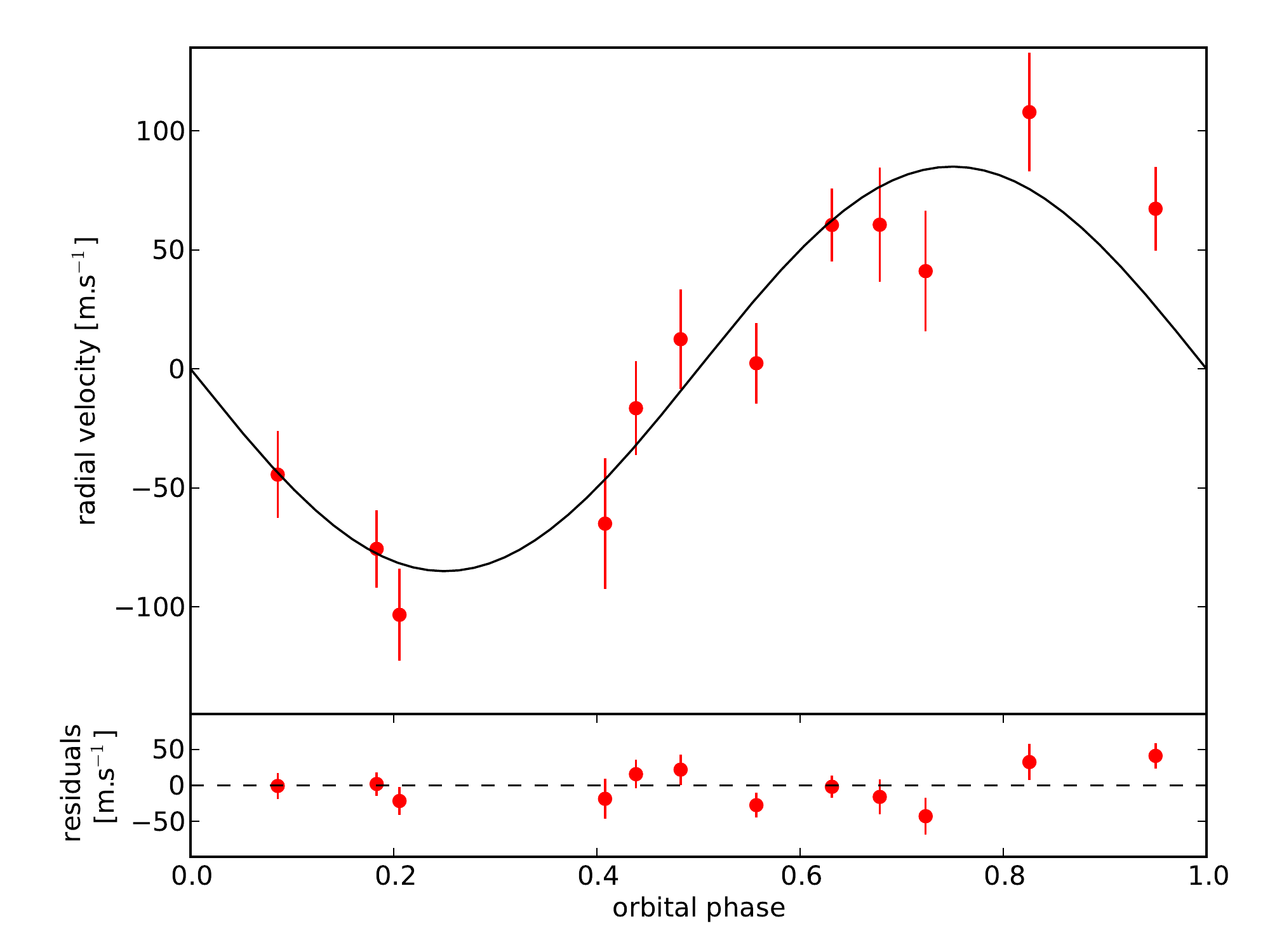}
\caption{Phase-folded radial velocity curve with our best-fit (top panel) and residuals (bottom panel).}
\label{196ph}
\end{center}
%\vspace{-0.5cm}
\end{figure}

\begin{figure}[h]
\begin{center}
\includegraphics[width=\columnwidth]{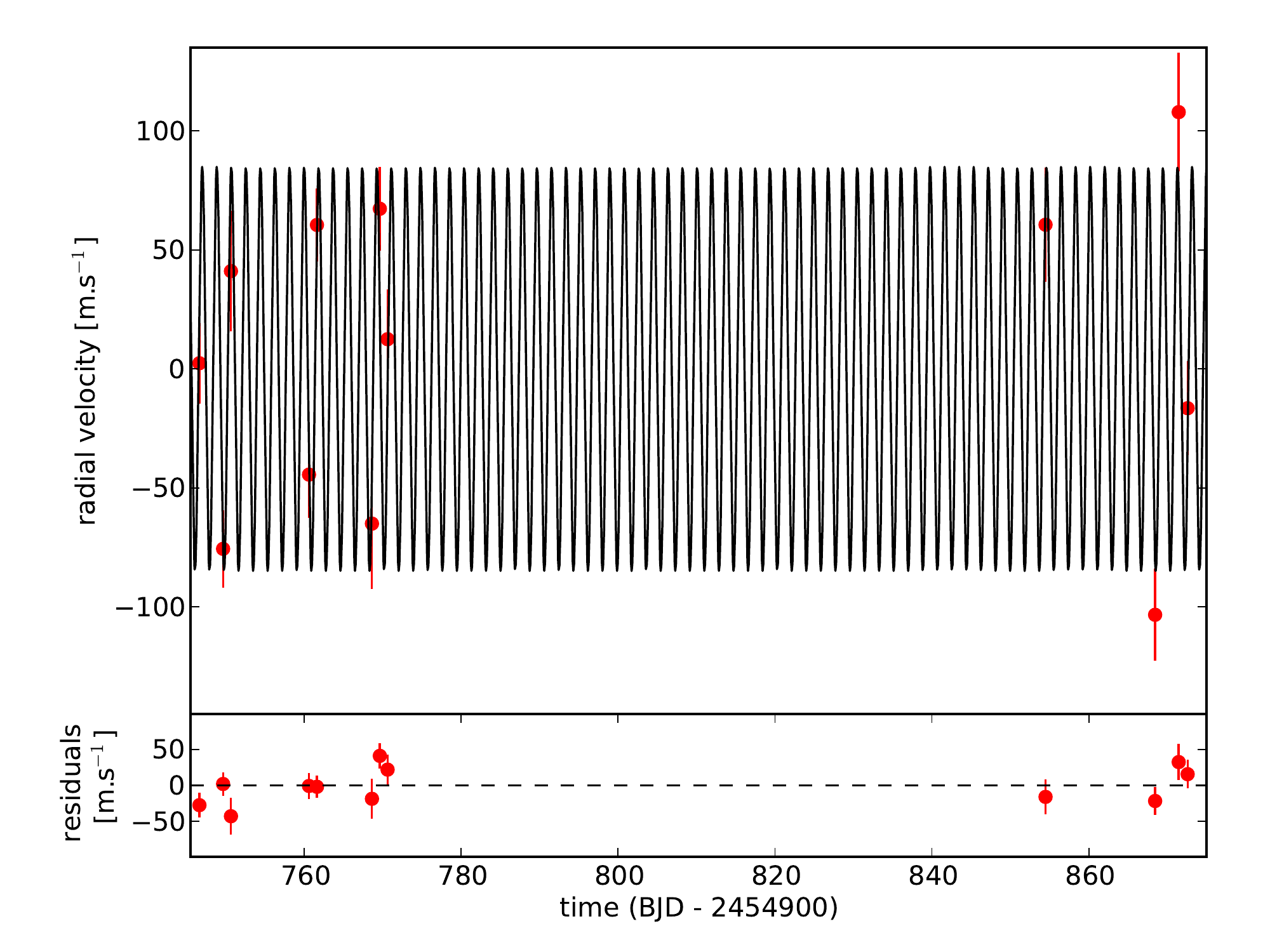}
\caption{Radial velocity curve with our best-fit (top panel) and residuals (bottom panel).}
\label{196time}
\end{center}
%\vspace{-0.5cm}
\end{figure}

These radial velocities are listed in Table \ref{196rv} and displayed in Fig. \ref{196ph} \& \ref{196time}. They show a clear variation compatible with the reflex motion of the parent star KOI-196 due to a planetary companion in phase with the \textit{Kepler} ephemeris. Since we found a secondary eclipse at phase $\sim$ 0.5 (c.f. Sect. \ref{sect.occultation}) and that circularization time-scales for such close-in planets are very short \citep{2007MNRAS.382.1768M}, we assumed the orbit to be circular and fitted a corresponding keplerian to the data. We found a best fit with a semi-amplitude $K = 85 \pm 11 \ms$ and a $\sigma_\mathrm{O-C}=24\ms$ which is comparable with the mean radial velocity uncertainty $\left<\sigma_\mathrm{rv} \right> = 20\ms$.\\

From the SOPHIE CCF parameters, one can estimate the \vsini and \met using equations described in \citet{2010A&A...523A..88B}. Using the FWHM and contrast of the CCF and assuming $\left(B - V \right) = 0.63$, we find a \vsini $= 4.1 \pm 1.0$ \kms~ and \met $ = -0.17 \pm 0.15$ dex.

\subsection{Blend analysis}

If the RV signal is mimicked by a diluted blended binary one can expect to see some correlation between the bisector span and radial velocities or different RV amplitude when processing the CCF with different spectral type mask templates \citep{2009IAUS..253..129B}. We first fitted the different RV datasets processed with K5 and F0 masks with a circular orbit model and found that $K_\mathrm{K5} = 84 \pm 11$ \ms, $\sigma_\mathrm{O-C} = 20\ms$ and $K_\mathrm{F0} = 85 \pm 14$ \ms, $\sigma_\mathrm{O-C} = 32\ms$ which is in good agreement with the amplitude derived with the G2 mask.\\

\begin{figure}[h]
\begin{center}
\includegraphics[width=\columnwidth]{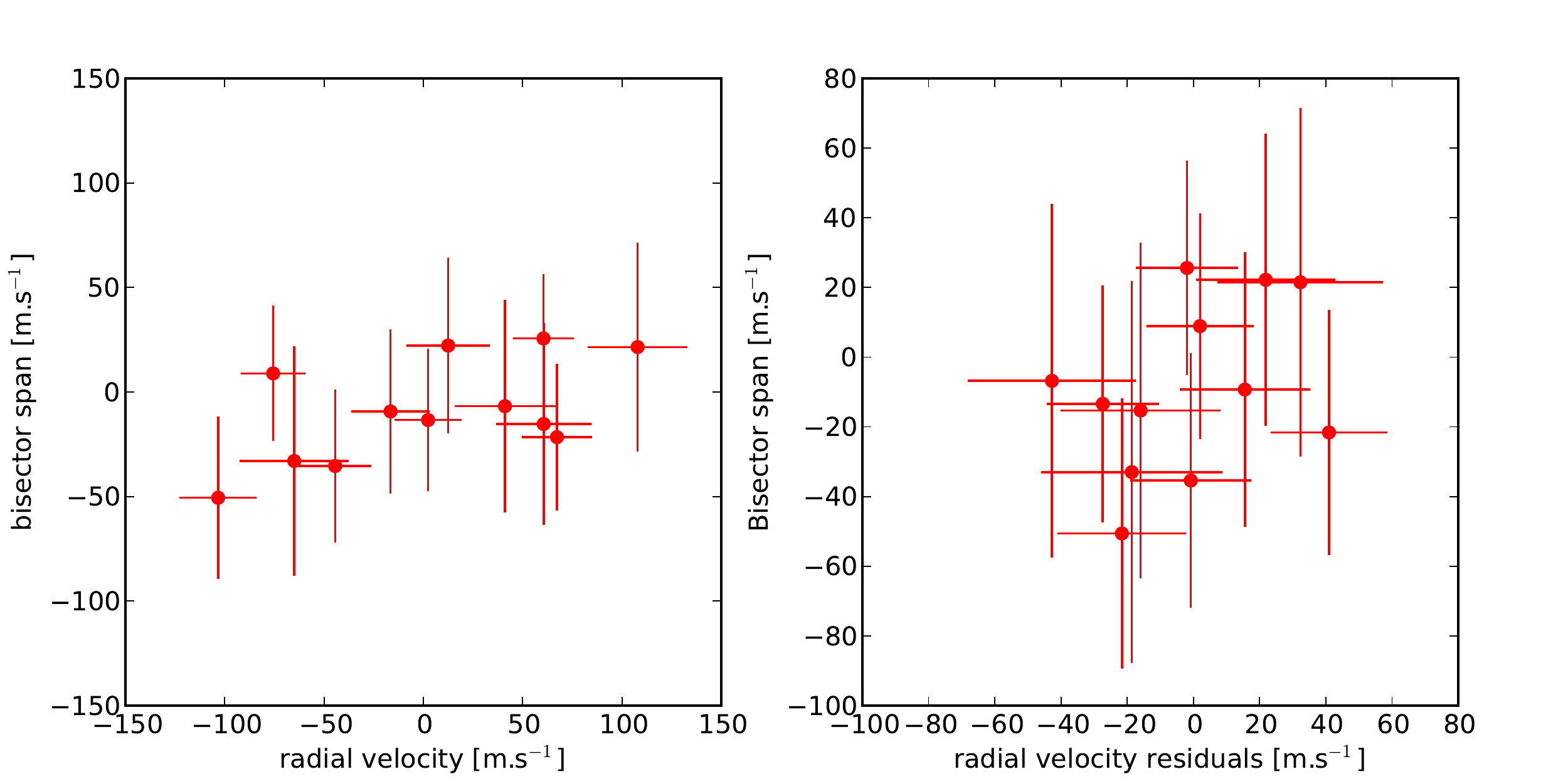}
\caption{CCF bisector as function of radial velocity (left panel) and radial velocity residuals (right panel).}
\label{196bis}
\end{center}
\end{figure}

Bisector span is also a key CCF diagnostics which is very sensitive to stellar activity and additional blended stellar component \citep[][Santerne et al., in prep.]{2001A&A...379..279Q, 2009IAUS..253..129B}. In order to assess the possibility that the radial velocity variations are not caused by a blended binary, we measured the bisector spans which are listed in Table \ref{196rv} and plotted in Fig. \ref{196bis}. They do not reveal any significant variation within 1-$\sigma$. Moreover, neither radial velocities, nor their residuals present a significant correlation with the bisector. These two checks allow us to secure the planetary nature of KOI-196b.\\

%\section{Spectral classification}
%\vspace{-0.4cm}

\section{System parameters}
\label{spclass}

To perform the spectral analysis of the KOI-196b host star, we used the four SOPHIE spectra took before June 2011 that were not affected by Moon reflected light at the time of the observations. Since such individual spectra have too low signal-to-noise ratio to allow a proper analysis, we co-added them after correcting of the barycentric Earth radial velocity projected through the line of sight and of the star's one. The resulting co-added spectrum has a signal-to-noise of 46 per element of resolution at thecontinuum at 5550 \AA~with a spectral resolution of $\sim 39\,000$. We used the semi-automatic software package VWA \citep{2002A&A...389..345B, 2008A&A...478..487B} to derive the stellar photospheric parameters as described in \citet{2010arXiv1005.3208B}.\\

Using the same methodology as presented in \citet{2010arXiv1005.3208B} we derived the projected rotational velocity $\vsini = 4.5 \pm 1.5$ \kms~ from the analysis of a few isolated spectral lines. This value is in good agreement with the \vsini estimated from the CCF assumed that the true value of the \vsini could be even lower since we are here limited by the spectral resolution of SOPHIE. For the atmospheric parameters we found \teff $= 5660 \pm 100 $ K,  \met $= -0.09 \pm 0.16$ dex and  \logg $ = 4.35 \pm 0.2$. It is known that the \logg~ is the most uncertain parameter derived from spectral analysis. The quite low signal-to-noise of our co-added spectrum combined with the moderate spectral resolution do not permit us to estimate this parameter more accurately. The derived stellar parameters are reported in Table \ref{starplanetparam}. They are in a good agreement within the error bars, with the estimates published in \citet{2011arXiv1102.0541B} \teff = 5585 K and \logg=4.51 with uncertainties up to 30\%.\\

From infrared photometry J and K available in the 2MASS archive, we computed the photometric \teff = 6000 $\pm$ 150 K following \citet{2010A&A...512A..54C}, assuming \met of -0.10 dex and taking a reddening of E(J-K)=0.059 \citep{1989ApJ...345..245C} into account. This photometric estimation is at 2-$\sigma$ from the one determined by spectral analysis and might be explained by an upper value of the reddening provided in the MAST database.\\

Finally, we used the stellar density derived by the transit modeling combined with the \teff from the spectral analysis to determine the star's fundamental parameters in the (\teff, \mr) space. The position of the star in the H.-R. diagram was compared to STAREVOL evolution tracks \citep{2010ApJ...715.1539T} minimizing the $\chi^{2}$ as described in \citet{2011A&A...528A..63S}. We found a main-sequence solution with $M_{\star} = 0.94 \pm 0.09$ \Msun and $R_{\star} = 0.966 \pm 0.032$ \Rsun. Main-sequence solutions are consistent with the absence of Lithium in the spectrum of the star and the low \vsini value we derived. We note that the inferred gravity surface \logg $= 4.44 \pm 0.10~$ is in good agreement with the spectroscopic value within 1-$\sigma$ as well as the one published by \citet{2011arXiv1102.0541B}: \logg=4.5. This solution led to an age of $ 7.7 \pm 3.4  $ Gyr.\\

These stellar parameters yield limb darkening coefficients in the \textit{Kepler} bandpass of $u_+= 0.66 \pm 0.03 $ and $u_-= 0.15  \pm 0.03$ \citep[using interpolated values from][]{2010A&A...510A..21S}. They are in good agreement with those obtained from the transit modeling within 2-$\sigma$ and 1-$\sigma$ for $u_+$ and $u_-$, respectively. \\

From the adopted stellar fundamental parameters, combined with results from the transit modeling and the analysis of the radial velocity observations, we find a mass and a radius of the planet of $M_{p}=0.49 \pm 0.09$ \Mjup and $R_{p} = 0.841 \pm 0.032$ \Rjup. The inferred mean density of the planet is $\rho_{p} =  1.02 \pm 0.18 \;g.cm^{-3}$.\\

\begin{table}[h]
%\vspace{1cm}
\centering
\caption{Star and planet parameters.}            
%\vspace{1cm}
\begin{minipage}{9cm} 
\setlength{\tabcolsep}{1.5mm}
\renewcommand{\footnoterule}{}                          
\begin{tabular}{l l}        
\hline\hline                 
\multicolumn{2}{l}{\emph{Ephemeris}} \\
\hline
Planet orbital period $P$ [days] & $1.855558 \pm 7.10^{-6} $  \\ 
Transit epoch $T_{tr}$ [BJD - 2454900] & $ 70.1803 \pm 0.0003 $  \\  
&\\
\multicolumn{2}{l}{\emph{Results from radial velocity observations}} \\
\hline    
Orbital eccentricity $e$  & 0 (fixed) \\
Semi-amplitude $K$ [\ms] & 85 $\pm$ 11 \\
Systemic velocity  $V_{r}$ [\kms] & -27.066 $\pm$ 0.007 \\
O-C residuals [\ms] & 24 \\
&\\
\multicolumn{2}{l}{\emph{Fitted transit parameters}} \\
\hline
Radius ratio $k=R_{p}/R_{\star}$ & $0.0895 \pm 0.0019$ \\ % 
Orbital inclination $I$ [deg] & $88.3 \pm 0.7$ \\  
transit duration $T_{14}$ [h] & $2.376 \pm 0.048$\\
limb darkening coefficient u$_{+}$ & 0.98$ \pm$ 0.17\\
limb darkening coefficient u$_{-}$ & 0.16 $\pm$ 0.20 \\
&\\
\multicolumn{2}{l}{\emph{Deduced transit parameters}} \\
\hline
Scaled semi-major axis $a/R_{\star}$ & $6.43 \pm 0.05$ \\ % 
Impact parameter $b$ &  $0.19 \pm 0.07$ \\ %
%phase of ingress / egress $\theta_{1}$ & $0.02599 \pm 9.10^{-5}$\\
\mr [solar units]& $1.012 \pm 0.009 $\\ %
%linear limb darkening coefficient u$_{a}$ & 0.57$ \pm$ 0.18\\
%quadratic limb darkening coefficient u$_{b}$ & $ 0.41 \pm 0.18$ \\
Stellar density $\rho_{\star}$ [$g\;cm^{-3}$] & $1.46 \pm 0.04 $\\  
&\\
\multicolumn{2}{l}{\emph{Fitted and deduced Out-of-Transit parameters}} \\
\hline
phase of secondary transit $\phi_\mathrm{occ}$ & $0.499 ^{_{+0.004}}_{^{-0.011}}$\\
planet day-side to stellar flux ratio $R_\mathrm{day}$ [ppm] & $64 ^{_{+10}}_{^{-12}} $\\
planet night-side to day-side flux ratio $F_\mathrm{N/D}$ & $ < 0.24 $\\
relative stellar brightness $z_\mathrm{lev}$ & $ 0.99993 \pm 0.00001$\\
$e \cos \omega $ & $ -0.002 ^{_{+0.006}}_{^{-0.017}}$\\
&\\
\multicolumn{2}{l}{\emph{Spectroscopic parameters }} \\
\hline
Effective temperature \teff[K] & 5660 $\pm$ 100\\ 
Metallicity \met [dex]&  $-0.09 \pm 0.16  $ \\   % 
Stellar rotational velocity {\vsini} [\kms] & 4.5 $\pm$ 1.5\\   %
Spectral type & G2V \\
&\\
\multicolumn{2}{l}{\emph{Stellar physical parameters from combined analysis}} \\
\hline
Star mass $M_\star$ [\Msun] &  $0.94 \pm 0.09 $\\ 
Star radius $R_\star$[\Rsun] & $0.966 \pm 0.032$ \\   
Surface gravity log\,$g$$^a$ [dex]& 4.44 $\pm$ 0.10 \\ %
Age of the star $t$ [Gy] & $ 7.7 \pm 3.4 $ \\
Distance of the system [pc] & 730 $\pm$ 70 \\  
&\\
\multicolumn{2}{l}{\emph{Planetary physical parameters from combined analysis}} \\
\hline
Orbital semi-major axis $a^b$ [AU] & $ 0.029 \pm 0.001$\\ % 
Planet mass $M_{p}$ [\Mjup] &  $0.49 \pm 0.09 $ \\ % 
Planet radius $R_{p}$[\Rjup]  & 0.841 $\pm$ 0.032  \\ %
Planet density $\rho_{p}$ [$g\;cm^{-3}$] & $1.02 \pm 0.18$\\ 
Geometric albedo $A_\mathrm{g}$ &  $0.30\pm0.08$ \\
Day-side temperature $^c$  $T_\mathrm{day}$ [K] &$ 1930 \pm 80 $ \\  %  
&\\       
\hline\hline
\vspace{-0.5cm}
\footnotetext[1]{derived from M$_\star$ and R$_\star$.}\footnotetext[2]{derived from a/R$_\star$ and R$_\star$.}
\footnotetext[3]{considering no atmospheric thermal circulation.} 
\end{tabular}
\end{minipage}
\label{starplanetparam}   
%\vspace{-0.5cm}
\end{table}

\section{Geometric albedo characterization}
\label{albedocharac}

\label{albedovsthermal}
The high accuracy of the \textit{Kepler} photometry as well as the large number of orbital periods covered during Q1 \& Q2 allowed us to clearly detect both the secondary transit and the phase variation in the optical (a 5-$\sigma$ detection, see Sect. \ref{sect.occultation}). Assuming a pure reflecting planet without thermal emission we found a geometric albedo of $A_{g} = 0.33 ^{_{+0.07}}_{^{-0.08}}$ in the \textit{Kepler} bandpass \citep{2006ApJ...646.1241R}. Even in the optical, part of the observed light from the planet could be due to thermal emission \citep{2007ApJ...667L.191L, 2009Natur.459..543S}. The observed occultation depth is thus a combination of the reflected light and the thermal emission:

\begin{equation}
\delta_\mathrm{occ}=\frac{F_{p_{th}}}{F_{\star}} + \frac{F_{p_{ref}}}{F_{\star}}
\end{equation}
where $F_{p_{th}}$ is the thermal flux generated by the planet and $F_{p_{ref}}$ is the reflected light. Using eq. 14 in \citet{2006ApJ...646.1241R} the reflected component of the occultation depth expressed as:

\begin{equation}
\frac{F_{p_{ref}}}{F_{\star}}=A_\mathrm{g} \left(\frac{R_{p}}{a}\right)^{2}
\end{equation}
for which $R_{p}$/$a$ is well constrained by the modeling of the primary transit.\\

To estimate the fraction of thermal emission that contributes to the planetary occultation, we computed the ratio between the flux of the star and the expected thermal flux of the planet assuming a blackbody emission. If the planet is tidally locked, its day-side temperature can be calculated using the following expression \citep{2011ApJ...729...54C}: 

\begin{equation}
T_\mathrm{d}=T_\mathrm{eff}\sqrt{\frac{R_{\star}}{a}}\left(1-A_{B}\right)^{\frac{1}{4}}\left[\frac{2}{3}-\frac{5}{12}\varepsilon\right]^{\frac{1}{4}}
\label{Teq}
\end{equation}

where the Bond albedo $A_{B}$ is fixed to zero assuming blackbody emission and $\varepsilon$ is the circulation efficiency that is allowed to vary from $\varepsilon = 1$ for a full redistribution of the energy budget from the day-side towards the night-side to $\varepsilon = 0$ for no heat circulation in the atmosphere.\\

We used the \citet{2004astro.ph..5087C} atmospheric model\footnote{http://www.stsci.edu/hst/observatory/cdbs/castelli\_kurucz\_atlas.html}, $S^\mathrm{CK}_{\lambda}$, of a solar-like star to estimate the stellar flux and integrated over the \textit{Kepler} bandpass $\Omega_{\lambda}$\footnote{http://keplergo.arc.nasa.gov/CalibrationResponse.shtml}. Then, we integrated the Planck function at the day-side brightness temperature $T_\mathrm{d}$ of the planet over the \textit{Kepler} bandpass. The occultation depth due to thermal emission is the ratio between the two fluxes, multiplied by the surface areas of the planet and the star, respectively :

\begin{equation}
\frac{F_{p_{th}}}{F_{\star}} = \pi k^{2}\frac{\int_{\lambda}\frac{2hc^{2}}{\lambda^{5}}\left[exp\left(\frac{hc}{k_{B}\lambda T_\mathrm{d}}\right)-1\right]^{-1}\Omega_{\lambda}\;\mathrm{d}\lambda}   {\int_{\lambda}S^\mathrm{CK}_{\lambda}\Omega_{\lambda}\;\mathrm{d}\lambda}
\end{equation}
where $h$ is the Planck constant, $k_{B}$, the Boltzmann constant and $c$ the speed of light in the vacuum.\\

\begin{figure}[h]
\begin{center}
\includegraphics[width=\columnwidth]{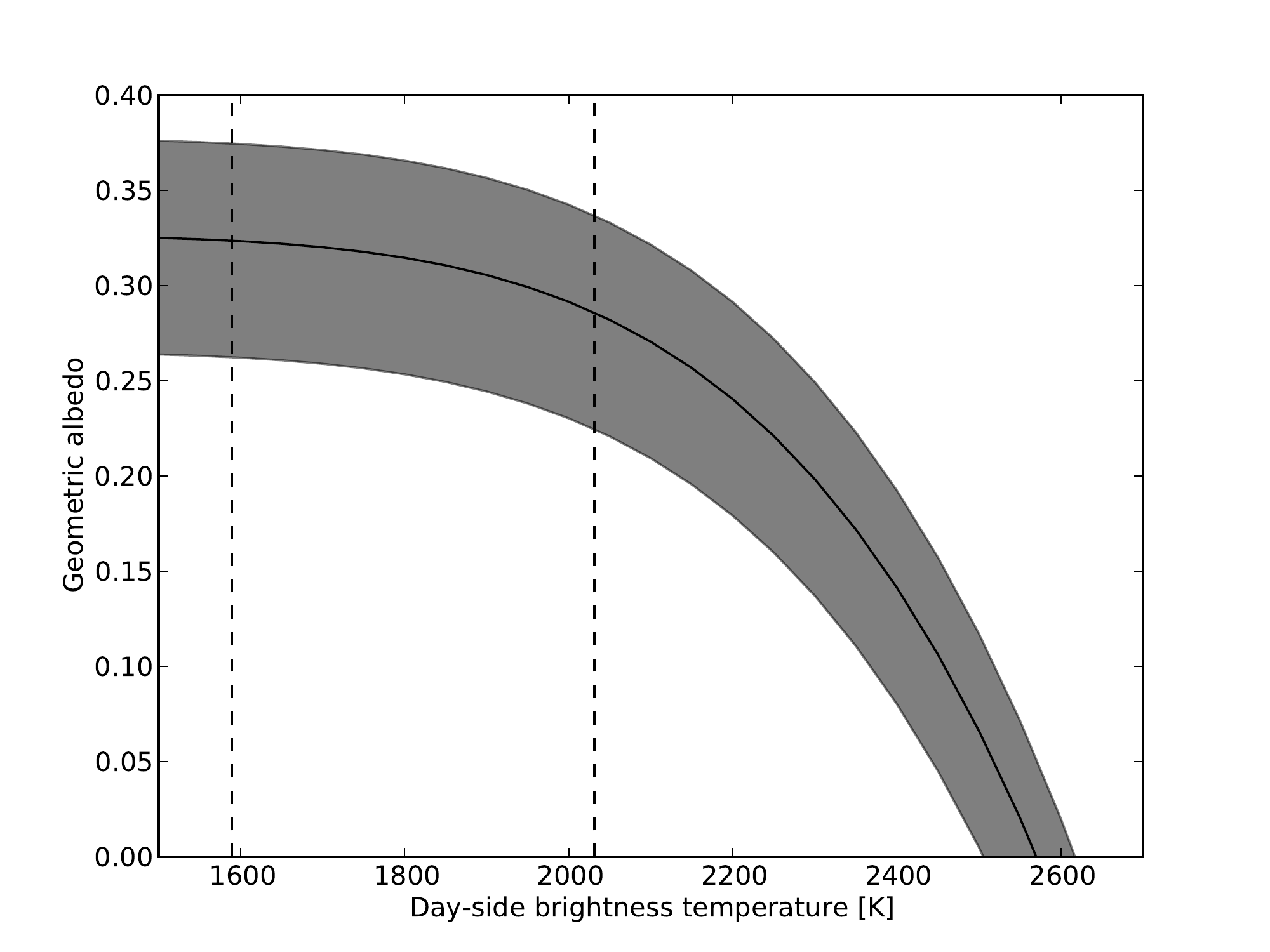}
\caption{Geometric albedo as function of planet day-side brightness temperature. Dashed lines represent the brightness temperature of the planet assuming zero Bond albedo and a perfect heat redistribution , $\varepsilon=1$ ; T=1590 $\pm$ 60 K (left) and no thermal circulation in the atmosphere $\varepsilon=0$ ; T=2030 $\pm$ 60 K (right). The grey band covers the albedo values allowed by the 1-$\sigma$ uncertainty on the occultation depth which is assumed to be the dominant uncertainty.}
\label{albedo}
\end{center}
\end{figure}

Fig. \ref{albedo} displays the geometric albedo in the \textit{Kepler} bandpass as a function of the brightness temperature of the planet as derived from the previous equations. We found that the thermal emission of the planet is negligible within the \textit{Kepler} bandpass and we found a geometric albedo of $A_{g} = 0.3\pm0.08$. Using this range of values for the albedo and assuming that the geometric albedo related to the Bond albedo through the relation: $A_{B}=2/3A_{g}$ for a perfect Lambertian sphere \citep[e.g.][]{2006ApJ...646.1241R}, and assuming that the energy redistribution is not efficient ($\varepsilon = 0$) for such a strongly irradiated planet, we estimated the day-side temperature of the planet of $T_\mathrm{day} =1930 \pm 80 $K.\\

%Using the 1-$\sigma$ upper limit on the night-side to day-side flux ratio compared with the expected thermal emission of the night-side, we can put an upper limit on the night-side brightness temperature of $T_{B}^\mathrm{night} < 2180 K$ as shown in Fig. \ref{night}. It is not expected that the night-side hemisphere of a such planet is hotter than the day-side. The quite low constrain on the night-side to day-side flux ration $F_{N/D}$ do not permit to put any constrain on the night-side temperature. Even with a precision of a 1 or 2 ppm as expected for PLATO, it will not possible to constrain this temperature less than the day-side equilibrium temperature. To better estimate this propriety, one must observe the thermal phase variation with infrared space telescopes like Spitzer or the upcoming JWST.\\

%\begin{figure}[h]
%\begin{center}
%\includegraphics[width=\columnwidth]{night_temperature.pdf}
%\caption{Expected thermal to stellar flux ratio of planet night-side as function of the planet night-side brightness temperature assuming a blackbody emission. The dashed-line is the 1-$\sigma$ upper limit value on the observed night-side to day-side flux ratio which correspond to a night-side brightness temperature of 2180K.}
%\label{night}
%\end{center}
%\end{figure}

\section{Discussion}
\label{discussion}

\subsection{A non-inflated low-mass hot-Jupiter}

\begin{figure}[h]
\begin{center}
\includegraphics[width=\columnwidth]{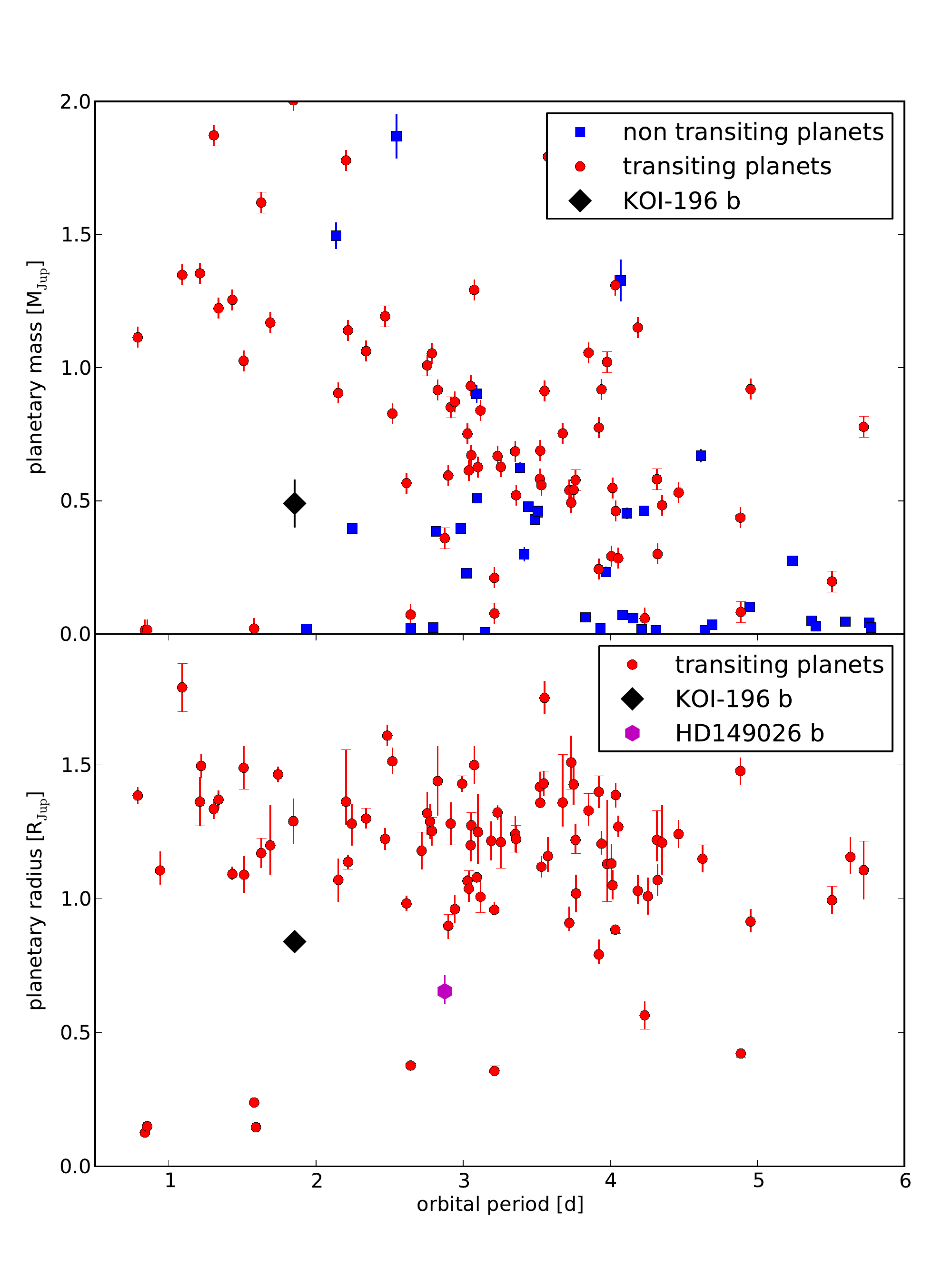}
\caption{Planetary mass (upper panel) and radius (lower pannel) as a function of orbital period for exoplanets with orbital period less than 6 days \cite[source:][]{2011PASP..123..412W}. Transiting exoplanets are the red circles and RV planets are the blue squares. We note that for non-transiting planets, the minimum $M\sin i$ is represented here. KOI-196b is marked  by a black diamond and HD149026b is marked with a magenta hexagon.}
\label{dzone}
\vspace{-0.5cm}
\end{center}
\end{figure}

With an orbital period of $1.855558 \pm 7.10^{-6} $ d, a radius of 0.841 $\pm$ 0.032 \Rjup and a mass of  $0.49 \pm 0.09 $ \Mjup, KOI-196b is one of the rare objects with a very short orbital period ($P < 2$ d) and radius and mass smaller than Jupiter. Fig. \ref{dzone} shows KOI-196 compared with other planets in terms of orbital period, planetary radius and mass. Only HD212301b \citep{2006A&A...451..345L} exhibits similar characteristics in terms of mass and orbital period. However as the orbital inclination of this planet is still unknown, the true mass of HD212301b could be likely much higher. Using tables in \citet{2007ApJ...659.1661F}, we estimated that the core mass of KOI-196b possesses between 50 M$_{\oplus}$ and 100 M$_{\oplus}$ of heavy elements. This means that between $\sim$ 1/3 and $\sim$ 2/3 of KOI-196b's mass is concentrated in its core. Another high-density short-period planet with radius smaller than Jupiter's one is HD149026b \citep{2005ApJ...633..465S}. Out of the 141 transiting planets discovered so far, only KOI-196b and HD149026b are located in the hot-Jupiter desert with orbital period shorter than 3 days and with radius or mass smaller than Jupiter. This desert could not be explained by an observational bias since their short orbital periods make their detection easier. These two objects might share the same formation and/or evolution process \citep{2007MNRAS.376L..62B}.\\

\begin{figure}[h]
\begin{center}
\includegraphics[width=\columnwidth]{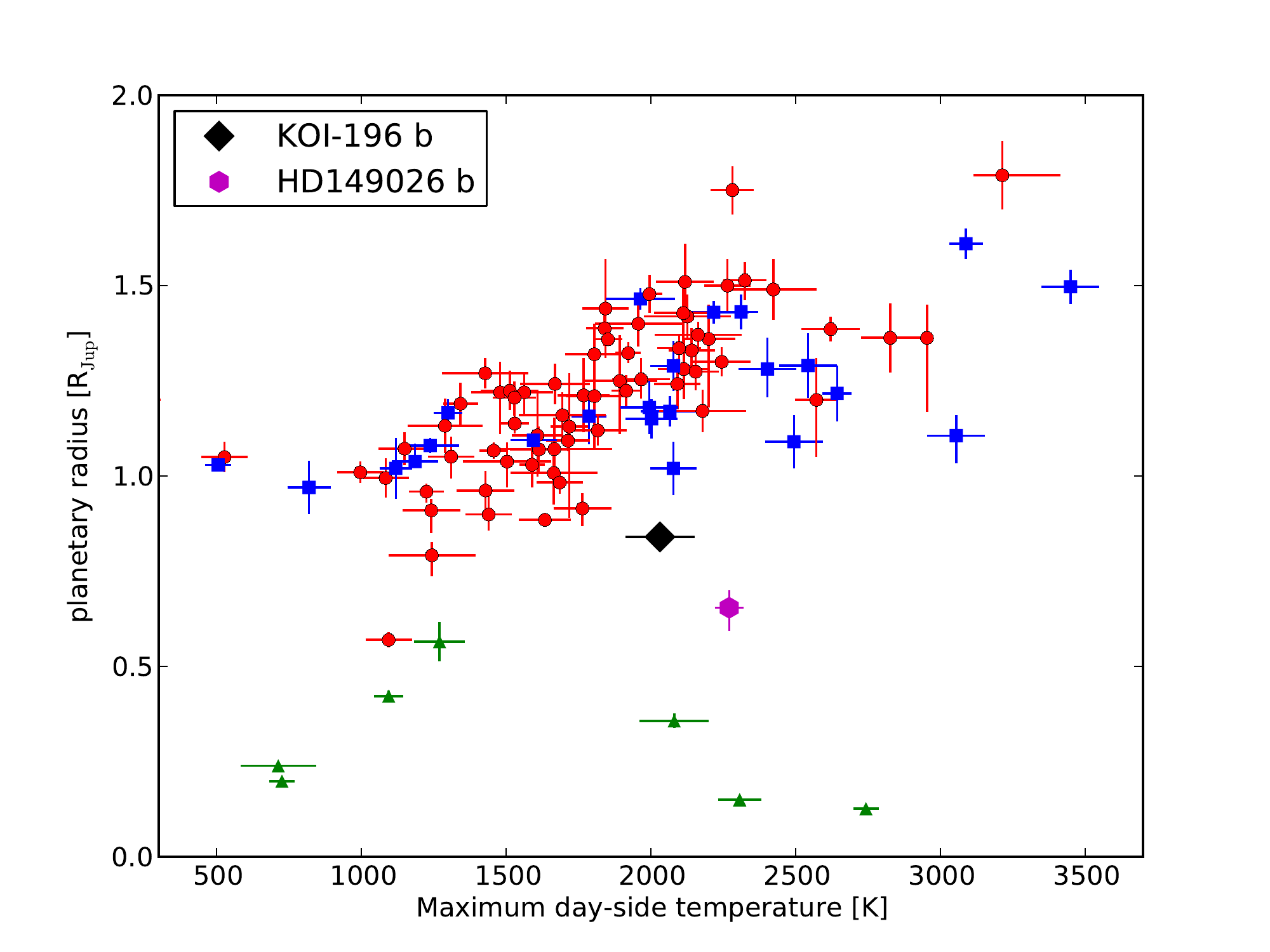}
\caption{Planetary radius vs estimated maximum day-side temperature assuming no heat circulation for all transiting planet \citep[source:][]{2011PASP..123..412W}. Transiting planets with mass $M_{p} > 2$ \Mjup are the blue squares. Transiting planets with mass in between $M_{p} = 0.1$ \Mjup and $M_{p} = 2$ \Mjup are the red circles. Transiting hot-Neptunes and Super-Earths are the green triangles. KOI-196b is marked by a black diamond and HD149026b is marked with a magenta hexagon.}
\label{Teqmax}
\vspace{-0.5cm}
\end{center}
\end{figure}

Fig. \ref{Teqmax} displays the planetary radius as function of the maximum day-side temperature assuming no heat circulation (i.e. $T_{\mathrm{day}}^\mathrm{max}=T_\mathrm{eff} / \sqrt{a/R_{\star}}$). There is a clear correlation between the radius and the expected maximum day-side temperature for hot-Jupiters less massive than 2 \Mjup as  expected theoretically \citep{2007ApJ...661..502B, 2008ApJ...687.1191L}. Once again, in the Saturn-Jupiter domain, KOI-196b and HD149026b appear to be outliers. This might be explained by an unusual high Bond albedo in comparison with other transiting planets as it is observed for KOI-196b. Indeed, it is expected that planets with a higher albedo are cooler (see eq. \ref{Teq}) and thus, are less inflated (see Fig. \ref{Teqmax}).\\

High mass-loss can explain this unusual low mass and low radius for a such short-period hot-Jupiter \citep{2011A&A...529A.136E}. Using eq. 15 and 22 in \citet{2007A&A...461.1185L} and assuming a efficient extraction of the gas by the extreme ultraviolet incident flux $\eta =1$, we estimate a current escape rate for KOI-196b of $\sim$ 0.0005 \Mjup / Gyr. Thus, mass-loss is a very unlikely process to explain the low mass and small radius of KOI-196b.\\

Further studies of internal structure, taking the high albedo of KOI-196b into account, are required to explain the small radius for this close-in planet. 

\subsection{A high albedo for a hot-Jupiter}

KOI-196b is the third hot Jupiter, discovered so far with a such high albedo. The first ones were \textit{Kepler}-7b \citep{2010ApJ...713L.140L} with $A_{g} = 0.32 \pm 0.03$ \citep{2011ApJ...730...50K, 2011ApJ...735L..12D} and HAT-P-7b \citep{2008ApJ...680.1450P} with an albedo of $A_{g} = 0.58 \pm 0.05$ \citep{2010ApJ...713L.145W}. Other giant planets with an observed occultation have a very low albedo in the optical $A_{B} \lesssim 0.1$ \citep{2011ApJ...729...54C}. Theoretical studies previously predicted the low-albedo of close-in planets \citep[e.g.][]{2000ApJ...538..885S} due to the high opacity of Na and K at optical wavelengths. \citet{2011ApJ...735L..12D} suggested that a depletion of such elements in the high atmosphere of Kepler-7b compared to solar abundances can explain its high albedo. \citet{2000ApJ...538..885S} have shown that a high albedo can also be explained by the presence of silicate layer high in the atmosphere. Finally, synthetic spectra of the thermal emission of hot-Jupiters from \citet{2008ApJ...678.1419F} suggests that thermal emission could be significantly different from a blackbody emission in the optical. In that case, contribution of the thermal light in the \textit{Kepler} bandpass can be significantly higher, leading to a lower value of the geometric albedo of KOI-196b. Infrared observations of the secondary eclipse and, if possible, the thermal phase variation of KOI-196b are needed to constrain the day-side and night-side temperature as well as to constrain unambiguously its geometric albedo and heat circulation \citep{2011ApJ...729...54C}.

\section{Conclusion}

Thanks to publicly-available \textit{Kepler} photometry and new high-resolution SOPHIE spectroscopic observations, we established the planetary nature of the Kepler Object of Interest KOI-196.01, now called ``KOI-196b'', one of the 1235 \textit{Kepler}'s candidates published by \citet{2011arXiv1102.0541B}. KOI-196b is a $0.49 \pm 0.09$ \Mjup -- $0.841 \pm 0.032$ \Rjup hot-Jupiter orbiting a slightly metal-poor G2V star in a $1.855558 \pmÊ0.000007 $d period. This planet is one of the rare close-in hot-Jupiters with a radius and mass lower than Jupiter suggesting that it is a non-inflated planet. We note that only 6 \textit{Kepler} candidates with priority 2, including KOI-196b, are expected to be hot-Jupiters with an orbital period shorter than 3 days and radius in between 0.5 \Rjup and 0.9 \Rjup : KOI-102.01, KOI-183.01, KOI-356.01, KOI-801.01 and KOI-883.01. Focusing follow-up efforts on them can help to specify the boundary of the short-period low mass and small radius hot-Jupiter desert, if exists.\\

Using the high-quality and long time-series of the \textit{Kepler} data, we detected the occultation as well as the optical phase variation of KOI-196b, orbiting a quite faint star (m$_\mathrm{V}Ê\sim $14.6). From the occultation depth and assuming a blackbody planetary thermal emission in the \textit{Kepler} bandpass, we estimated a geometric albedo of the planet to be $0.30\pm0.08$, which is markedly higher than the geometric albedos observed for most of the known hot-Jupiters \citep{2011ApJ...729...54C}. This leads to a day-side temperature of $T_\mathrm{day} =1930 \pm 80 $ K assuming no thermal redistribution in the atmosphere. Lower values for the albedo are still possible but would indicate a significant contribution of the planetary thermal emission in the optical. This can be unambiguously confirmed with infrared observations of the planetary occultation.\\

This detection demonstrates once again the efficiency of SOPHIE, a dedicated instrumentation on a 2-m class telescope for the ground-based follow-up of space mission like \textit{Kepler}, \textit{CoRoT} or the ESA M-class mission \textit{PLATO}, if selected.

\begin{table}[]
\centering
\setlength{\tabcolsep}{1.0mm}
\begin{minipage}[t]{7.5cm} 
\caption{SOPHIE measurements of KOI-196.}
\begin{tabular}{ccccc}
\hline
\hline
BJD & RV & $\pm 1\sigma_\mathrm{rv}$ & $V_\mathrm{span}$ & S/N/pix \\
(-2 454 900) & [\kms] & [\kms] & [\kms] & @550nm \\
\hline
 %& & & & \\
746.63646 & -27.063 & 0.017 & -0.013 & 16.8\\
749.65408 & -27.141 & 0.016 & 0.009 & 17.0\\
750.65676 & -27.024 & 0.025 & -0.007 & 14.7\\
760.60669 & -27.110 & 0.018 & -0.035 & 17.5\\
761.61891 & -27.005 & 0.015 & 0.026 & 19.1\\
768.62681 & -27.130 & 0.027 & -0.033 & 14.3\\
769.63221 & -26.998 & 0.018 & -0.022 & 19.3\\
770.62035 & -27.053 & 0.021 & 0.022 & 16.6\\
854.48407 & -27.005 & 0.024 & -0.015 & 13.7\\
868.45141 & -27.169 & 0.019 & -0.051 & 16.8\\
871.45738 & -26.957 & 0.025 & 0.021 & 17.4\\
872.59440 & -27.082 & 0.020 & -0.009 & 17.9\\
% & & & & \\

\hline
\hline
\end{tabular}
%\vspace{-0.3cm}
\label{196rv}
\end{minipage}
\end{table}

\begin{acknowledgements}
We thank the technical team at the Observatoire de Haute-Provence for their support with the SOPHIE instrument and the 1.93-m telescope and in particular the essential work of the night assistants. We are grateful to the \textit{Kepler} Team for giving public access to the corrected \textit{Kepler} light curve and for publishing a list of good planetary candidates to follow-up. Financial support for the SOPHIE observations from the ÒProgramme National de Plan\'etologieÓ (PNP) of CNRS/INSU, France is gratefully acknowledged. We also acknowledge support from the French National Research Agency (ANR-08-JCJC-0102-01). A.S. is grateful to Cilia Damiani for all the fruitful discussions. A.S.B. is supported by CNES.
\end{acknowledgements}

\end{document}